\documentclass[reprint,aps,pre]{revtex4-2}

\usepackage{graphicx}
\usepackage{caption}
\usepackage{subcaption}
\usepackage{float}
\usepackage{xcolor}
\usepackage{amssymb,amsmath,amsthm}
\usepackage[colorlinks,
            pdffitwindow=false,
            plainpages=false,
            pdfpagelabels=true,
            pdfpagemode=UseOutlines,
            pdfpagelayout=SinglePage,
            bookmarks=false,
            colorlinks=true,
            hyperfootnotes=false,
            linkcolor=blue,
            urlcolor=blue!50!black,
            citecolor=green!50!black]
{hyperref}
\usepackage{comment}
\usepackage{bm}

\graphicspath{{figs/}}

\newcommand{\revision}[1]{{#1}}
\newcommand{\secondrevision}[1]{{#1}}

\newcommand{\R}{\mathbb{R}}
\newcommand{\E}{\mathbb{E}}
\renewcommand{\P}{\mathbb{P}}
\newcommand{\dif}{\text{d}}
\DeclareMathOperator*{\argmin}{argmin}

\newcommand{\numA}{N}  
\newcommand{\numO}{M}  
\newcommand{\bap}{m}  
\newcommand{\x}{x}  
\newcommand{\oip}{a}  
\newcommand{\ois}{b}  
\newcommand{\oit}{c}  
\newcommand{\oiq}{d}  
\newcommand{\nip}{i}  
\newcommand{\nis}{j}  
\newcommand{\nit}{k}  
\newcommand{\sh}{c}  
\newcommand{\shv}{\boldsymbol{c}}  
\newcommand{\wsh}{\bar{c}}  
\newcommand{\wshv}{\bar{\boldsymbol{c}}}  
\newcommand{\wshmfe}{\wsh_{\text{\textsf{\tiny mMFE}}}}
\newcommand{\wshmfev}{\wshv_{\text{\textsf{\tiny mMFE}}}}
\newcommand{\ifac}{\alpha}  
\newcommand{\dcor}{\delta_{\text{cor}}}  
\newcommand{\dint}{\delta_{\text{int}}}  
\newcommand{\ind}{\mathbf{1}}  
\newcommand{\indx}[2]{\ind_{\{\x_{#1}=#2\}}}
\newcommand{\ev}{\boldsymbol{e}} 

\begin{document}
\title{Accurate mean-field equation for voter model dynamics on scale-free networks}

\author{Marvin Lücke}
\affiliation{Zuse Institute Berlin, Germany}

\author{Stefanie Winkelmann}
\affiliation{Zuse Institute Berlin, Germany}

\author{Péter Koltai}
\affiliation{Department of Mathematics, University of Bayreuth, Germany}

\begin{abstract}
	Understanding the emergent macroscopic behavior of dynamical systems on networks is a crucial but challenging task.
	One of the simplest and most effective methods to construct a reduced macroscopic model is given by mean-field theory.
	The resulting approximations perform well on dense and homogeneous networks but poorly on scale-free networks, which, however, are more realistic in many applications.
	In this paper, we introduce a modified version of the mean-field approximation for voter model dynamics on scale-free networks.
    The two main deviations from classical theory are that we use degree-weighted shares as coarse variables and that we introduce a correlation factor that can be interpreted as slowing down dynamics induced by interactions.  We observe that\secondrevision{, for moderate noise and comparable interaction rates,} the correlation factor is only a property of the network and not of the state or of parameters of the process.
	This approach achieves a significantly smaller approximation error than standard methods without increasing dimensionality.
\end{abstract}
\maketitle

\section{Introduction}
The voter model \cite{Clifford1973,Holley1975} and its many extensions (see, e.g., \cite{Castellano2009,Porter2016,Redner2019} for overviews) are among the most popular models for studying spreading or infection processes on networks.
In this paper, we focus on a general variant, the \emph{continuous-time noisy voter model} (CNVM) \cite{Luecke2023,Luecke2024a,Carro2016}.
Given a simple graph $G$ on $\numA$ nodes, each node \revision{$\nip \in \{0, \dots, \numA - 1\} =: [\numA]$} has a state $\x_\nip \in \{0, \dots, \numO - 1\} =: [\numO]$.
As the voter model is often applied to opinion dynamics, we refer to $\x_\nip$ as the \emph{opinion} of node $\nip$.
Each opinion $\x_\nip$ evolves over time according to a continuous-time Markov chain: if $\x_\nip$ is in state $\oip \in [\numO]$, it transitions to a different state $\ois \in [\numO]$ at the rate
\begin{equation} \label{eq:cnvm_rates}
	r_{\oip, \ois} \frac{d_{\nip, \ois}}{d_\nip} + \tilde{r}_{\oip, \ois},
\end{equation}
where $d_{\nip, \ois}$ denotes the number of neighbors of node $\nip$ with opinion $\ois$, $d_\nip$ is the degree
of node $\nip$, and $r_{\oip, \ois}, \tilde{r}_{\oip, \ois} \geq 0$ are model parameters.
The first term in~\eqref{eq:cnvm_rates} describes imitation: the more neighbors of node $\nip$ have opinion $\ois$, the higher the transition rate to $\ois$.
The second term $\tilde{r}_{\oip, \ois}$ is independent of the neighborhood and induces explorative behavior, which can also be interpreted as noise.

An important observable for such discrete-state systems is the vector of \emph{shares} \revision{$\shv =(\sh_\oip)_{\oip \in [\numO]} \in [0,1]^\numO$}, $\lVert \shv \rVert_1 = 1$, where $\sh_\oip$ describes the \revision{expected} fraction of nodes $i$ that have opinion $x_i = \oip$,
\revision{
i.e.,
\begin{equation}
	\sh_\oip :=
	\frac{1}{\numA} \sum_{\nip\in[\numA]}  \E[\indx{\nip}{\oip}],
\end{equation}
where $\indx{\nip}{\oip} \in \{0,1\}$ denotes the indicator function of the event $\x_\nip = \oip$.
}
Using mean-field theory \cite{Barrat2008}, it has been shown that, in the large-population limit $\numA \to \infty$, the trajectory $\shv(t)$ of time-dependent shares converges to the solution of a \emph{mean-field equation} (MFE), provided that the underlying network is sufficiently homogeneous and dense \cite{Luecke2023, Keliger2024, Gleeson2012, Ayi2021}.
For scale-free networks---e.g., generated by the Barabási--Albert (BA) model \cite{Albert2002} with preferential attachment parameter $\bap$---however, this classical mean-field approximation performs poorly \cite{Fernley2022,Gleeson2013,PastorSatorras2001}.
The main contribution of this paper is the derivation and validation of a \emph{modified mean-field equation} (mMFE) that can accurately capture the macroscopic behavior of the CNVM on \revision{scale-free} networks.

It is well known \cite{Luecke2023} that for the CNVM the MFE is a simple system of ordinary differential equations
\begin{equation} \label{eq:standard_mfe}
	\frac{\dif}{\dif t} \shv(t) = \sum_{\substack{\oip,\ois \in [\numO] \\ \oip \neq \ois}} \sh_\oip(t) \big(r_{\oip, \ois}\, \sh_{\ois}(t) + \tilde{r}_{\oip, \ois}\big) (\ev_\ois - \ev_\oip),
\end{equation}
where $\ev_\oip$ denotes the $\oip$-th unit vector in $\R^\numO$.
To achieve an accurate approximation for \revision{scale-free} networks, we introduce two modifications.
First, the imitation \revision{terms $r_{\oip, \ois} \sh_{\oip} \sh_{\ois}$}
are multiplied by a constant \emph{correlation factor} $0 < \ifac \leq 1$; second, the shares $\shv \in [0,1]^\numO$ are replaced by the \emph{degree-weighted shares} \revision{$\wshv = (\wsh_\oip)_{\oip \in [\numO]}\in [0,1]^\numO$},
\revision{
\begin{equation}
	\wsh_\oip :=
	\frac{1}{\sum_{\nip\in[\numA]} d_\nip} \sum_{\nip\in[\numA]} d_\nip  \E[\indx{\nip}{\oip}].
\end{equation}
}
The dynamics of these degree-weighted shares is then approximated by the mMFE
\begin{equation} \label{eq:modified_mfe}
	\frac{\dif}{\dif t} \wshv(t) = \sum_{\substack{\oip,\ois \in [\numO] \\ \oip \neq \ois}} \wsh_\oip(t) \big(\ifac\, r_{\oip, \ois}\, \wsh_{\ois}(t) + \tilde{r}_{\oip, \ois}\big) (\ev_\ois - \ev_\oip).
\end{equation}
The significance of the degree-weighted shares is already known in some contexts \cite{Bianconi2002, Suchecki2005}, and they were specifically determined to be excellent \emph{collective variables} for the CNVM on BA networks using a data-driven method developed by the authors~\cite{Luecke2024}.

In this paper, we derive the above mMFE~\eqref{eq:modified_mfe}, motivating why these modifications are necessary for \revision{scale-free} networks, and show how to obtain the correlation factor~$\ifac$.
We analyze the quality and robustness of the mMFE depending on the model parameters $r$ and $\tilde{r}$, using a theoretical error analysis as well as numerical studies.
We show that the mMFE approximates the degree-weighted shares exceptionally well when the imitation rates $r_{\oip, \ois}$ are similarly large and there is moderate noise~$\tilde{r}$.
Moreover, in this setting the mMFE is robust in the sense that $\ifac$ does not depend on the specific model parameters and initial state\revision{---hence, the correlation factor appears to be a universal attribute of the network, more precisely of the network's statistical properties}.
In other parameter regimes, the approximation quality degrades slightly.
\revision{
Due to its popularity, the numerical examples presented in the main text utilize the BA model to generate scale-free networks, but almost identical results can be obtained for other examples like the \textit{configuration model} \cite{Bollobas1980}, as shown in Appendix~\ref{appendix:configuration-model}. 
}
Specifically for BA networks, another interesting result is that the correlation factor $\ifac$ increases towards $1$ with increasing preferential attachment parameter $\bap$ of the BA network.
\revision{
Finally, we show that the mMFE outperforms standard methods of similar complexity like the standard homogeneous MFE \eqref{eq:standard_mfe} and pair-approximation (PA) \cite{Porter2016,Gleeson2013} significantly.
While higher-dimensional methods like the \emph{heterogeneous} MFE and PA \cite{Porter2016, Pugliese2009}, \emph{approximate master equation} \cite{gleeson2011high}, \emph{quenched} MFE and PA \cite{Mata2013}, or \emph{dynamic cavity method} \cite{Neri2009} are able to resolve more detailed information about the system than the mMFE, they also require substantially more variables.
The mMFE is, on the other hand, able to provide a parsimonious macroscopic description of the collective dynamics using only a minimal set of variables: the degree-weighted shares~$\wshv$.
}

Accompanying code for reproducing the numerical experiments presented in this paper is provided at \cite{lucke_2025_17122500}.

\section{The modified mean-field equation}
In this section, we derive the modified mean-field equation (mMFE) from the evolution equation of the first-order moments.
We have to rely on two assumptions, the \emph{correlation assumption} and the \emph{interchangeability assumption}, which are explained in more detail later.
For simplicity, we consider the case of $\numO = 2$ opinions.
The general case is analogous, and is treated in Appendix~\ref{sec:appendix}. In the binary-state case, it is sufficient to only work with the (expected) weighted share $\wsh_1$ of opinion $1$, which we will denote as $\wsh = \wsh_1 \in [0,1]$ for simplicity.
It can be defined as
\begin{equation} \label{def:wsh}
	\wsh := \E\bigg[\frac{1}{\sum_\nip d_\nip} \sum_{\nip\in[\numA]} d_\nip \x_\nip \bigg] = \frac{1}{\sum_\nip d_\nip} \sum_{\nip\in[\numA]} d_\nip  \E[\x_\nip],
\end{equation}
where expectation is taken with respect to randomness in the dynamical evolution and possibly also in the initial state. Recall that in the CNVM the rate at which node $i$ in state $x_i = 0$ transitions to state $1$ is given by
\begin{equation}
	r_{0,1} \frac{1}{d_\nip} \sum_{ \nis \in [\numA] }  A_{\nip, \nis} \x_\nis + \tilde{r}_{0, 1},
\end{equation}
where \revision{$(A_{i,j})_{i,j\in [N]}\in\{0,1\}^{N\times N} $} denotes the adjacency matrix of the network.
Hence, the evolution of the first-order moment $\E[\x_\nip]$ is
\begin{align}
	 & \frac{\dif}{\dif t} \E[\x_\nip] =                                                                                                         \nonumber \\
	 & \P(\x_\nip = 0) \ \E\Big[\frac{r_{0,1} }{d_\nip} \sum_\nis A_{\nip, \nis} \x_\nis + \tilde{r}_{0, 1} \,\Big\vert\, \x_\nip = 0 \Big] \ -  \nonumber \\
	 & \P(\x_\nip = 1) \ \E\Big[\frac{r_{1,0}}{d_\nip} \sum_\nis A_{\nip, \nis} (1 - \x_\nis) + \tilde{r}_{1, 0} \,\Big\vert\, \x_\nip = 1 \Big].
\end{align}
By inserting the equality $\P(\x_\nip = 0)\, \E[\x_\nis \mid \x_\nip = 0] = \E[(1 - \x_\nip) \x_\nis]$, this implies
\begin{align} \label{eq:E}
	\frac{\dif}{\dif t} \E[\x_\nip] & =  \frac{r_{0,1}}{d_\nip} \sum_\nis A_{\nip, \nis} \E[(1 - \x_\nip) \x_\nis] + \tilde{r}_{0, 1} \E[1 - \x_\nip] \nonumber \\
	                                & -  \frac{r_{1,0}}{d_\nip} \sum_\nis A_{\nip, \nis} \E[\x_\nip (1 - \x_\nis)] - \tilde{r}_{1, 0} \E[\x_\nip].
\end{align}
Hence, it follows that
\begin{align} \label{eq:first_order_moment}
	 & \frac{\dif}{\dif t} \wsh = \frac{1}{\sum_\nip d_\nip} \sum_\nip d_\nip \frac{\dif}{\dif t} \E[\x_\nip] =                                                         \\
	 & \frac{1}{\sum_\nip d_\nip} \Big( r_{0,1} \sum_{\nip,\nis} A_{\nip, \nis} \E[(1 - \x_\nip) \x_\nis] + \tilde{r}_{0, 1} \sum_\nip d_\nip \E[1 - \x_\nip] \nonumber \\
	 & \quad - r_{1,0} \sum_{\nip,\nis} A_{\nip, \nis} \E[\x_\nip (1 - \x_\nis)] - \tilde{r}_{1, 0} \sum_\nip d_\nip \E[\x_\nip] \Big).
\end{align}

We break up the second-order moments via the \emph{(weak) correlation assumption}
\begin{equation} \label{eq:assume_cor}
	\sum_{\nip,\nis} A_{\nip, \nis} \E[(1 - \x_\nip) \x_\nis] \overset{!}{=}  \sum_{\nip,\nis} A_{\nip, \nis}\, \ifac\, \E[1 - \x_\nip] \E[\x_\nis]
\end{equation}
for some constant \textit{correlation factor} $0 < \ifac < 1$.
This implies that, on average, it is more likely to find two neighbors in the same state than if they were stochastically independent \revision{(in which case $\alpha =1$)} \footnote{This is because if $x_i$ and $x_j$ were stochastically independent, we would have that
\begin{align*}
\E[(1-x_i) x_j] &= \P[x_i=0,\ x_j=1] \\
    &= \P[x_i=0] \P[x_j=1] = \E[1-x_i] \E[x_j].
\end{align*}
Hence, the weak correlation assumption states that, on network-average, it is less probable than in the independent case that neighbors have different opinions.}.
Given this assumption, we have
\begin{align}
	 & \frac{\dif}{\dif t} \wsh = \frac{1}{\sum_\nip d_\nip} \Big(                                                                                           \nonumber \\
	 & \quad r_{0,1} \sum_{\nip,\nis} A_{\nip, \nis} \, \ifac\, \E[1 - \x_\nip] \E[\x_\nis] + \tilde{r}_{0, 1} \sum_\nip d_\nip \E[1 - \x_\nip] \nonumber              \\
	 & - r_{1,0} \sum_{\nip,\nis} A_{\nip, \nis} \, \ifac \, \E[\x_\nip] \E[1 - \x_\nis] - \tilde{r}_{1, 0} \sum_\nip d_\nip \E[\x_\nip] \Big).
\end{align}
Next, we apply the \emph{(weak) interchangeability assumption}
\begin{equation} \label{eq:assume_int}
	\sum_{\nip,\nis} A_{\nip, \nis}\, \E[1 - \x_\nip] \E[\x_\nis] \overset{!}{=} \sum_{\nip,\nis} A_{\nip, \nis}\, (1 - \wsh) \wsh,
\end{equation}
\revision{which states that the \emph{average} pairwise interaction between the expected states behaves as if all nodes shared a common state $\wsh$. 
This is substantially weaker than a ``strong'' interchangeability assumption of the form $\E[x_i] = \zeta$ for all nodes $i$, which would imply $c = \bar c = \zeta$.
}
Applying \eqref{eq:assume_int} yields the mMFE for $\numO = 2$ opinions
\begin{align} \label{eq:modified_mfe_M2}
	\frac{\dif}{\dif t} \wsh \ = & \quad \ifac \, r_{0,1} (1 - \wsh) \wsh + \tilde{r}_{0,1} (1 - \wsh) \nonumber \\
	                             & - \ifac \, r_{1,0}\, \wsh (1 - \wsh) - \tilde{r}_{1,0}\, \wsh.
\end{align}
An analogous derivation yields the mMFE for an arbitrary number $\numO$ of opinions, as defined in~\eqref{eq:modified_mfe}, see appendix \ref{sec:appendix} for details.

It is clear that, due to possible violation of the two assumptions, the solution of the mMFE \eqref{eq:modified_mfe} does not exactly match the actual weighted shares~$\wshv$.
To distinguish the two, we will continue to refer to the true weighted shares as $\wshv$ and denote the solution of the mMFE as~$\wshmfev$.

\subsection{Relation to classical mean-field theory}
The two assumptions \eqref{eq:assume_cor} and \eqref{eq:assume_int} are closely related to those used in classical mean-field theory, but adapted to reflect the properties of scale-free networks.
In the classical theory, it is assumed that all node degrees grow unbounded as the number of nodes increases and hence any two neighbors become stochastically independent, i.e., choosing $\ifac = 1$ in \eqref{eq:assume_cor}.
In scale-free networks, however, a large portion of nodes have finite low degree, regardless of how large the network grows.
As a result, the states of neighboring nodes do not become stochastically independent, but instead remain positively correlated.
The important observation here is that the average correlation between neighbors can, with some limitations outlined later, be well captured by a constant factor~$\ifac < 1$.

Moreover, in the classical theory it is typically argued that nodes become interchangeable in the mean-field limit, so that the expected state of every node is simply given by the average state, e.g., $\E[\x_\nip] = \sh$ for all $\nip$ in the binary-state case ($\numO = 2$).
We observe numerically that, for BA networks, this interchangeability is also valid on average; however, the expected state of each node is related to the \emph{degree-weighted} average state $\wsh$ instead, as described in assumption~\eqref{eq:assume_int}.
As mentioned earlier, this finding was inspired by the authors' work~\cite{Luecke2024}, which computationally revealed that the weighted shares are excellent \emph{collective variables} for the CNVM on BA networks.
\revision{This means that their future evolution depends only on the current value of $\wsh$, and on no further information from the full state~$\boldsymbol{x}=(\x_\nip)_{\nip\in [N]}$.}

Further verification of the above claims will be given in the following sections.

\subsection{Error analysis}
Again, for simplicity, we assume the case of $\numO = 2$ opinions.
We define the error terms $\dcor, \dint \in [-1, 1]$ of the two assumptions \eqref{eq:assume_cor} and \eqref{eq:assume_int} as
\begin{align}
	\dcor & := \frac{1}{\sum_\nip d_\nip} \sum_{\nip,\nis} A_{\nip, \nis} \Big( \label{eq:dcor}              \\
	      & \qquad \qquad  \ifac\, \E[(1 - \x_\nip)] \E[\x_\nis] - \E[(1 - \x_\nip) \x_\nis] \Big) \nonumber \\
	\dint & :=  \frac{1}{\sum_\nip d_\nip} \sum_{\nip,\nis} A_{\nip, \nis} \Big( \label{eq:dint}             \\
	      & \qquad \qquad (1 - \wsh) \wsh - \E[(1 - \x_\nip)] \E[\x_\nis] \Big). \nonumber
\end{align}
Note that $\dcor(t;\ifac)$ depends on time and on the factor $\ifac$, whereas $\dint(t)$ only depends on time.
Let $F$ denote the right-hand side of the mMFE \eqref{eq:modified_mfe_M2}, i.e.,
\begin{equation}
	\frac{\dif}{\dif t} \wshmfe =  F(\wshmfe).
\end{equation}
Recall that the evolution of the exact weighted shares $\wsh$ is given by the first-order moment equation \eqref{eq:first_order_moment}.
Inserting the error terms $\dcor$ and $\dint$ into \eqref{eq:first_order_moment} yields
\begin{equation}
	\frac{\dif}{\dif t} \wsh = F(\wsh) + (r_{1,0} - r_{0,1}) (\dcor + \ifac \, \dint).
\end{equation}
Moreover, note that $F$ is Lipschitz with constant $K(\ifac) := \ifac \lvert r_{1,0} - r_{0,1} \rvert + \lvert \tilde{r}_{1,0} - \tilde{r}_{0,1}\rvert$.
Thus, by Gronwall's inequality we obtain an upper bound for the error
\begin{align}
	 & \lvert \wsh(t) - \wshmfe(t; \ifac) \rvert \leq                                                                                    \\
	 & \quad e^{K(\ifac) t}  \int_0^t \big\lvert (r_{1,0} - r_{0,1}) (\dcor(s; \ifac) + \ifac \, \dint(s)) \big\rvert \dif s . \nonumber
\end{align}
Since $\lvert \dcor + \ifac \, \dint \rvert \leq 2$, this automatically implies a small error if $r_{1,0} \approx r_{0,1}$ and $\tilde{r}_{1,0} \approx \tilde{r}_{0,1}$.
However, if these rate differences are large, the overall error can also become large, depending on $\dcor$ and $\dint$---that is, on how well the two assumptions \eqref{eq:assume_cor} and \eqref{eq:assume_int} of the derivation hold.

Let the number of opinions $\numO$ be arbitrary again.
To assess the quality of the mMFE, we employ the normalized $L_1$-error over an interval $[0, T]$
\begin{equation} \label{eq:L1_error}
	L_1(T; \ifac) := \frac{1}{T} \int_0^T \frac{1}{2} \lVert \wshv(t) - \wshmfev(t) \rVert_1 \dif t.
\end{equation}
(Note that $\frac{1}{2} \lVert \wshv(t) - \wshmfev(t) \rVert_1 \in [0, 1]$ is the \emph{total variation distance} between discrete probability distributions \cite{Rachev1991}.)
Then, choosing the factor
\begin{equation} \label{eq:optimal_alpha}
	\ifac^* := \argmin_\ifac L_1(T; \ifac)
\end{equation}
yields the optimal fit of $\wshmfev$ to $\wshv$ on the interval $[0, T]$.

\subsection{Parameter regimes}

\begin{figure*}
	\centering
	\includegraphics[width=.9\textwidth]{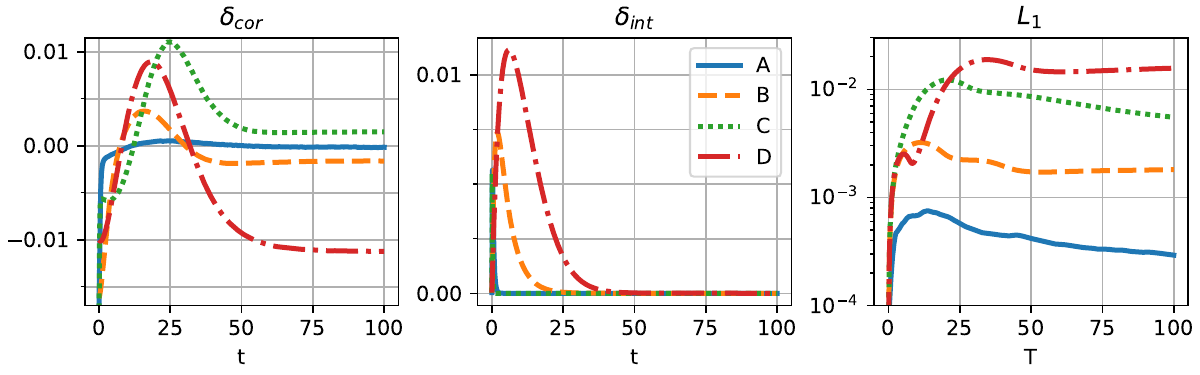}
	\vspace{-.3cm}
	\caption{Errors $\dcor(t; \ifac^*)$, $\dint(t)$  and $L_1(T; \ifac^*)$, as defined in \eqref{eq:dcor}, \eqref{eq:dint}, and \eqref{eq:L1_error}, respectively, in the example systems of the different regimes A,\smallskip B,\smallskip C,\smallskip D. Estimated from numerical simulations.}
	\label{fig:errors_assumptions}
\end{figure*}

Building on the above error analysis, we define four parameter regimes of the CNVM, ordered by ascending mMFE error:
\begin{enumerate}
	\item \textbf{Regime A}: Very similar rates $\big|\frac{r_{\oip, \ois}}{r_{\oit, \oiq}} -1\big| < 0.05$
	      for all valid combinations $a,b,c,d \in [M]$, and moderate (i.e., small but non-negligible) noise $\tilde{r}_{\oip, \ois}$.
	      Example: $r_{0,1} = 1$, $r_{1,0} = 0.98$, $\tilde{r}_{0,1}=\tilde{r}_{1,0}=0.01$.
	\item \textbf{Regime B}: Similar rates $\big|\frac{r_{\oip, \ois}}{r_{\oit, \oiq}} -1\big| < 0.5$ and moderate noise $\tilde{r}_{\oip, \ois}$.
	      Example: $r_{0,1} = 1$, $r_{1,0} = 0.7$, $\tilde{r}_{0,1}=\tilde{r}_{1,0}=0.01$.
	\item \textbf{Regime C}: Similar rates and (partially) no noise. Example: $r_{0,1} = 1$, $r_{1,0} = 0.98$, $\tilde{r}_{0,1}=\tilde{r}_{1,0}=0$.
	\item \textbf{Regime D}: Dissimilar rates and (partially) no noise. Example: $r_{0,1} = 1$, $r_{1,0} = 0$, $\tilde{r}_{0,1}= 0$, $\tilde{r}_{1,0}=0.3$ (i.e., SIS model \cite{Kiss2017}).
\end{enumerate}
(The case of large noise is not considered, as it would render the impact of the underlying network negligible.)
\revision{A possible intuitive interpretation is that regimes A and B describe opinion or strategy dynamics with bidirectional imitation and some exploration---regime A being nearly neutral, regime B allowing moderate bias. Regime C reflects pure imitation without exploration (e.g., consensus formation), while regime D captures strongly asymmetric, directional contagion (e.g., disease or innovation spreading).} 
\revision{Note that the particular values $0.05$ and $0.5$ that serve as the similarity thresholds for regimes A and B were chosen as useful rules of thumb but do not carry a special meaning.
In fact, the mMFE error grows steadily with the rate dissimilarity, see Fig. \ref{fig:error_vs_rates} in the appendix for details.
}

We outline the differences between these regimes using the above-mentioned example parameters (rescaled to exhibit comparable behavior over the interval $t \in [0,100]$).
In all regimes, we use the same initial state, in which a randomly chosen $10\%$ of nodes start with opinion 1 and the remainder with opinion 0,
and the same BA network with parameter $\bap=3$ and $\numA = 10^4$ nodes.
Fig.~\ref{fig:errors_assumptions} shows, for each regime, the errors $\dcor$ and $\dint$ for the optimal $\ifac^*$ as defined in \eqref{eq:optimal_alpha} with $T=100$, estimated from 15000 model simulations, as well as the resulting $L_1$-error~\eqref{eq:L1_error}.
Fig.~\ref{fig:trajs} (left) shows the corresponding trajectories $\wsh(t)$ and~$\wshmfe(t)$.

\begin{figure}
	\centering
	\includegraphics[width=\linewidth]{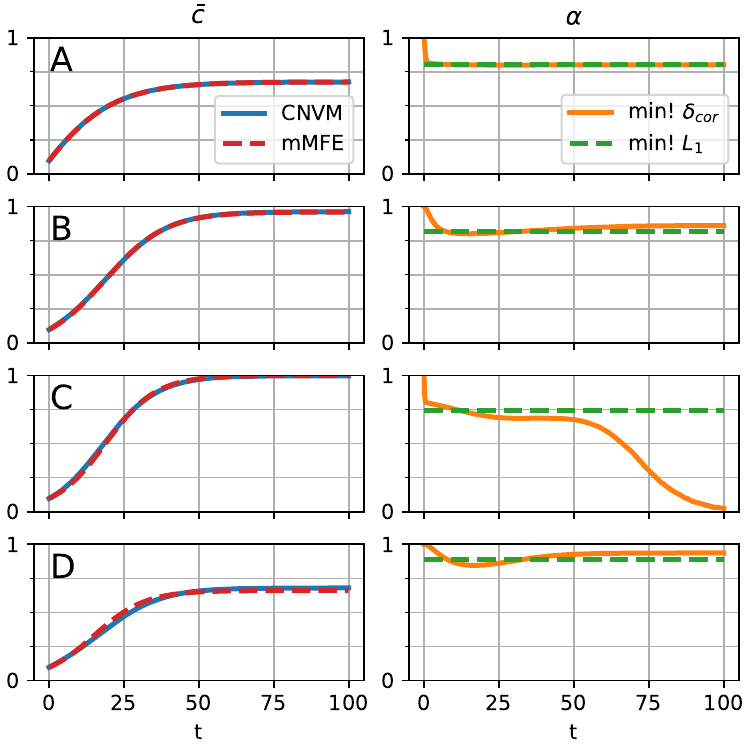}
	\caption{Trajectories of $\wsh$, \revision{given in~\eqref{def:wsh},} estimated from simulations, for the example systems of the different regimes A B,C,D (left), compared to the mMFE \eqref{eq:modified_mfe_M2}.
		The mMFE uses the optimal $\ifac^*$ as in \eqref{eq:optimal_alpha}, shown on the right (green dashed).
		The right plot also shows the $\ifac$ that minimizes $\dcor(t;\ifac)$ for each $t$ (orange solid).}
	\label{fig:trajs}
\end{figure}

Regime A exhibits the smallest $L_1$-error because $|r_{1,0} - r_{0,1}|$ is small and the error terms $\dcor$ and $\dint$ are close to $0$ for all $t$.
In regime B the error terms $\dcor$ and $\dint$ become slightly larger, and since $|r_{1,0} - r_{0,1}|$ also increases compared to regime A, the $L_1$-error is slightly larger.

A substantially different behavior appears in regime C.
Due to the absence of noise, clusters of identical opinion are highly stable, and the system eventually falls into an absorbing state with $\x_\nip = 0$ for all $\nip$ or $\x_\nip = 1$ for all $\nip$.
As a result, the expected correlation between neighbors increases over time, so that the chosen $\ifac^*$ is too small (i.e., $\dcor < 0$) for $t < 20$ but too large (i.e., $\dcor > 0$) for~$t > 20$.
Since $|\dcor|$ is rather large most of the time, the overall $L_1$-error is larger than in regimes A and~B.

Finally, in regime D (SIS model) the largest error occurs.
During the transient phase, up to approximately $t=30$, the infection ($\x_\nip = 1$) spreads from the initially infected nodes, forming clusters of infected nodes that create strong correlations between neighbors.
Over time, these clusters get broken up as nodes spontaneously become susceptible again, leading to an equilibration of $\wsh$ and a reduction in neighbor correlations.
As a result, $\dcor$ is much larger than $0$ in the transient phase and well below $0$ in equilibrium.
Combined with a comparatively large $\dint$ and $|r_{1,0} - r_{0,1}|$, the overall $L_1$-error is greater than in all other regimes.

The most important observation from this experiment is that in regimes A and B, when imitation rates are similar and noise is present, the mMFE \eqref{eq:modified_mfe} produces only a very small error.
In regimes C and D, on the other hand, the correlation between neighbors varies over time, leading to a larger error.
The correlation assumption \eqref{eq:assume_cor} is not sufficiently satisfied in these regimes for a time-constant $\ifac$, as Fig.~\ref{fig:trajs} (right) illustrates.

\begin{figure}
	\centering
	\includegraphics[width=\linewidth]{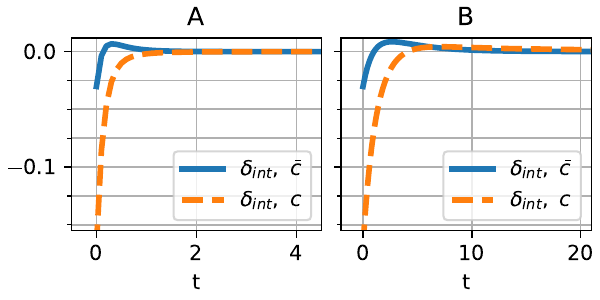}
	\vspace{-0.6cm}
	\caption{Interchangeability error $\dint$, see \eqref{eq:dint}, when using the regular shares $\sh$ instead of the degree-weighted shares $\wsh$, for regimes A and B (with different initial conditions than in Fig.~\ref{fig:errors_assumptions}).}
	\label{fig:interchange}
\end{figure}

It should also be noted that the use of the degree-weighted shares $\wsh$, as opposed to the regular shares $\sh$, is crucial for obtaining a small error in the initial phase.
Fig.~\ref{fig:interchange} illustrates that using the regular shares $\sh$ in the interchangeability assumption~\eqref{eq:assume_int} instead of $\wsh$ can result in an error $\dint$ that is magnitudes larger.
This is demonstrated using an initial state $\boldsymbol{x}$ where the top $10\%$ of the highest-degree nodes have opinion $1$ and the rest have opinion~$0$, resulting in a large discrepancy between $\sh$ and~$\wsh$.
\revision{This behavior reflects the disproportionate dynamical influence of high-degree nodes in the CNVM: hubs are imitated more often and affect many neighbors simultaneously, which strongly shapes the early and transient dynamics. The degree-weighted shares $\bar c$ encode this effect and therefore satisfy the closure assumptions much better than the unweighted shares $c$, as Fig.~\ref{fig:interchange} illustrates.}

\subsection{Practical considerations}
Computing the factor $\ifac^*$ by solving~\eqref{eq:optimal_alpha} can be computationally demanding, as many simulations may be required to estimate the weighted shares trajectory $\wshv(t)$, $t \in [0,T]$.
\revision{
In the ergodic system case, i.e., if $\tilde{r}_{\oip, \ois} > 0$ for all $\oip\neq\ois$, 
}
the factor $\ifac^*$ can be calculated efficiently using only the steady-state weighted shares \revision{$\wshv_\infty=(\wsh_{\infty,a})_{a\in [M]}$}, which can be obtained from a single (sufficiently long) simulation started from an arbitrary initial state.
Consider that in the steady-state of the mMFE \eqref{eq:modified_mfe} it is
\begin{align}
	 & 0 = \frac{d}{dt} \wsh_\oip =                                                                                                                                                            \\
	 & \ifac \sum_{\ois \neq \oip} \wsh_\oip \wsh_\ois (r_{\ois, \oip} - r_{\oip, \ois}) + \sum_{\ois \neq \oip} \wsh_\ois \tilde{r}_{\ois, \oip} - \wsh_\oip \tilde{r}_{\oip, \ois} \nonumber
\end{align}
for all $\oip \in [\numO]$, where the summation is performed over all $b$ except~$b=a$. This implies that
\begin{equation} \label{eq:ifac_efficient}
	\ifac = \frac{\sum_{\ois\neq \oip} (\wsh_\oip \tilde{r}_{\oip, \ois} - \wsh_\ois \tilde{r}_{\ois, \oip})}{\sum_{\ois\neq \oip} \wsh_\oip \wsh_\ois (r_{\ois, \oip} - r_{\oip, \ois})}.
\end{equation}
Hence, $\ifac$ is given by inserting $\wshv_\infty$ into \eqref{eq:ifac_efficient}.
\revision{
Note that, due to the ergodicity, it is $0 < \wsh_{\infty,\oip} < 1$ for all $\oip$, and hence the above expression is well-defined
}
and yields the optimal $\ifac^*$ as defined in \eqref{eq:optimal_alpha} with~$T \to \infty$.

The usefulness of the mMFE would be severely limited if $\ifac$ had to be calculated from \eqref{eq:optimal_alpha} or \eqref{eq:ifac_efficient} for every system of interest. However, the next section shows that this is not necessary, as the optimal $\ifac$ is robust across initial states and a wide range of system parameters.

\section{Robustness}
The previous section illustrated that, especially in regimes A and B, the mMFE with an appropriate choice of the correlation factor $\ifac$ yields a very good approximation.
This section shows that this choice of $\ifac$ is robust to variations in the initial condition and the parameters $r$ and~$\tilde{r}$.
Thus, the mMFE is \emph{general} in the sense that a suitable $\ifac$ provides a good approximation for any initial condition and any model parameters in regimes A and~B.
Moreover, the optimal $\ifac$ depends only on the preferential attachment parameter $\bap$ of the BA graph, as will be discussed later.

The robustness of $\ifac$ is verified with the following numerical study.
We consider a BA network with preferential attachment parameter $\bap = 3$ and $\numA = 10^5$ nodes, and the CNVM with $\numO = 3$ opinions.
We sample 200 initial states $\boldsymbol{x} \in [\numO]^\numA$ using four different strategies to obtain a diverse set of random initial conditions:
\begin{enumerate}
	\item \emph{Uniform}: The initial opinion of each node is drawn uniformly from $[\numO]$.
	\item \emph{Uniform shares}: Sample shares $\shv$ uniformly at random from the simplex $\{\shv \in [0,1]^\numO \mid \lVert \shv \rVert_1 = 1\}$, and assign opinions randomly so that $\shv$ is matched.
	\item \emph{By degree}: Again, sample shares $\shv$ uniformly. Choose a random order of opinions, assign the first opinion to the $\numA\sh_1$ nodes with largest degree, the second opinion to the next $\numA\sh_2$ remaining nodes with largest degree, and so on.
	\item \emph{Clusters}: 
	      Randomly assign opinions to a few nodes. Then, iteratively, take a random node that already has an opinion and let it pass this opinion to all of its neighbors that do not yet have one. Repeat this process until all nodes have an opinion.
\end{enumerate}
Moreover, for each of the 200 initial states, we sample random parameters from regimes A and B via
\begin{equation}
	r_{\oip, \ois} \sim \text{Uniform}([0.95, 1.05])
\end{equation}
and set $\tilde{r}_{\oip, \ois} = 0.01$ for all $\oip, \ois \in [\numO]$.
The correlation factor $\ifac$ is computed with \eqref{eq:ifac_efficient} from a single long simulation starting from a random one of the initial states, which yields
\begin{equation}
	\ifac = 0.81.
\end{equation}
(In fact, every initial state would result in approximately the same $\ifac$, as this experiment will show.)
For each of the 200 initial states and parameter sets, we computed the trajectories of the weighted shares $\wshv(t)$ via simulation and compared them to the mMFE solution $\wshmfev(t)$ under identical initial conditions and parameters.

The results of this experiment are shown in Fig.~\ref{fig:errors_robust}.
We observe errors below 0.005 for all trajectories, with a median error smaller than 0.003.
This confirms that setting $\ifac = 0.81$ yields a very small error for any choice of initial condition and parameter set in regimes A and B, thereby demonstrating the robustness of the mMFE.

\begin{figure}
	\centering
	\includegraphics[width=\linewidth]{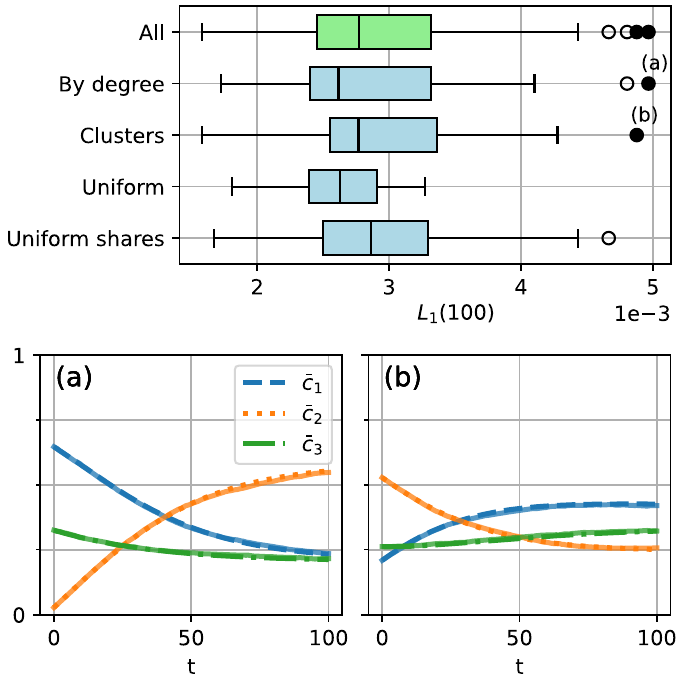}
	\vspace{-0.7cm}
	\caption{$L_1$-error \eqref{eq:L1_error} of the mMFE for different initial conditions and parameters in regime A and B (top). Even for the two trajectories (a) and (b) with largest errors (bottom) the approximation is very good. Solid lines indicate true weighted shares and dashed/dotted lines the solution of the mMFE.}
	\label{fig:errors_robust}
\end{figure}

As mentioned before, the proper choice of $\ifac$ depends only on the \revision{underlying network structure}, i.e., on the preferential attachment parameter $\bap$.
Since $\bap$ determines the minimum node degree, a larger $\bap$ generally reduces neighbor correlations.
Consequently, as $\bap$ increases, the optimal $\ifac$ also increases.
For very large $\bap$, neighbors are almost independent, implying $\ifac \approx 1$.
This behavior is illustrated in Fig. \ref{fig:pref_attach}, which shows the optimal $\ifac$ computed from \eqref{eq:ifac_efficient} for different values of~$\bap$.
\revision{
Moreover, note that the specific value of $\ifac$ also differs depending on the generation method of the networks, see appendix \ref{appendix:configuration-model} for a comparison between BA model and configuration model.
}

It should be noted that the above results---especially regarding the validity of the mMFE and the robustness of $\ifac$---are accurate only for sufficiently large networks.
In the examples discussed here, we found that $\numA \geq 10^4$ is a reasonable threshold.
For smaller $\numA$, the mMFE, like other mean-field approximations, loses accuracy, \revision{see Fig. \ref{fig:network_size} in the appendix.}

\begin{figure}
	\centering
	\includegraphics[width=.9\linewidth]{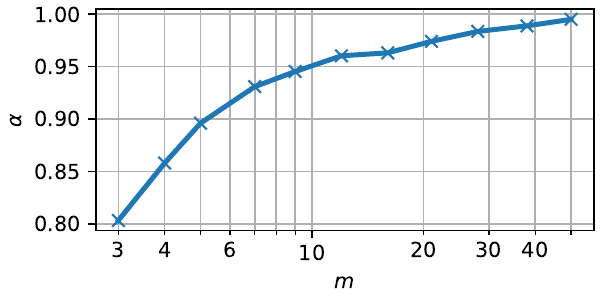}
	\vspace{-0.3cm}
	\caption{The factor $\ifac$ in the mMFE \eqref{eq:modified_mfe} depends on the preferential attachment parameter $m$ of the BA graph.}
	\label{fig:pref_attach}
\end{figure}

\section{Comparison to classical approximations}
\begin{figure}
	\centering
	\includegraphics[width=\linewidth]{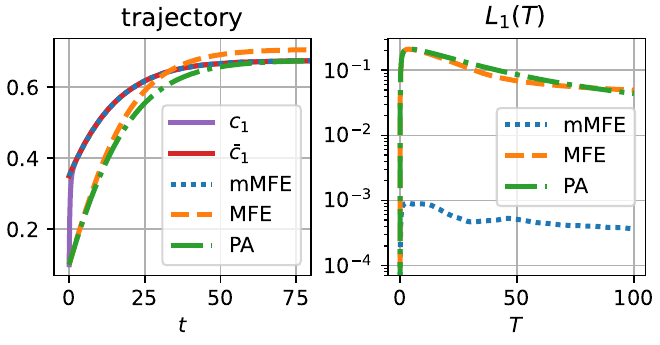}
	\caption{The shares $\sh_1$ and weighted shares $\wsh_1$ estimated from simulations, in comparison to the mMFE, MFE, and PA. Model parameters from regime A.
		The left plot shows the trajectories, while the right plot shows the $L_1$-error \eqref{eq:L1_error} between the mMFE and $\wshv$, as well as between MFE and $\shv$, and PA and $\shv$.}
	\label{fig:comparison}
\end{figure}

In this section we compare the mMFE, which is a $(\numO - 1)$-dimensional approximation, to other standard low-dimensional moment-closure approximations: the MFE, which also has dimension $(\numO - 1)$, and the \emph{pair-approximation} (PA), which has a dimension of $(\numO - 1) + \numO (\numO - 1)$.
The PA results from closing the moment equations at second order and therefore includes the number of edges between different opinions as additional variables compared to the MFE.
\revision{
A detailed derivation of the PA for the CNVM is omitted here for brevity but the general idea as well as the resulting equations are presented in appendix \ref{appendix:pair-approx}.
}
As Fig. \ref{fig:comparison} illustrates, the classical approximations (MFE and PA) fail to capture the system behavior on BA networks, whereas the mMFE incurs only a very small error.
This is not surprising, since the classical approximations assume a \emph{homogeneous} network structure.
The well-known \emph{heterogeneous} variants of the MFE and PA, or the \emph{approximate master equation}, are better suited for scale-free networks~\cite{gleeson2011high,Porter2016,Gleeson2013,Pugliese2009}.
However, as these resolve the opinion shares \emph{per degree}, they involve substantially more variables than the mMFE.
For instance, on a BA network with $\numA = 10^4$ nodes and \revision{preferential attachment parameter} $\bap = 4$, there are on average about 100 different node degrees, implying that the heterogeneous MFE has 100 times more variables than our mMFE.
\revision{The \emph{quenched} MFE \cite{Mata2013} even resolves the expected state \emph{per node} such that the required number of variables is $\numA$ times larger than that of the mMFE.
}
Hence, to compare reduced order models with the same degree of efficiency, we only report comparison between the mMFE and the homogeneous MFE and PA.

\revision{Needless to say, the above mentioned methods and similar approaches, such as the heterogeneous MFE or approximate master equation, resolve (and hence can potentially predict) more detailed information about the dynamics. In contrast, our results indicate that the degree-weighted shares together with the mMFE provide a parsimonious macroscopic description that captures the relevant collective behavior of the CNVM on certain scale-free networks. The robustness and resulting universality of the parameter~$\ifac$ make the mMFE a lightweight and computationally efficient surrogate when only a macroscopic quantity is of interest.
}

\section{Conclusion}
In this paper, we derived the mMFE~\eqref{eq:modified_mfe}, a modified version of the classical mean-field equation (MFE) that describes the dynamics in terms of weighted shares and includes an average ``correlation factor''~$\ifac$. We then verified that this equation provides an accurate and  efficient approximation of the continuous-time noisy voter model on scale-free networks generated by the Barabási--Albert or the configuration model.
It outperforms the standard MFE and pair-approximation (PA), while remaining low-dimensional.
We found that the approximation error depends on the underlying parameter regime; in particular, in regimes C and D---corresponding to (partially) noise-free dynamics and/or very dissimilar imitation rates---a time-constant correlation factor $\ifac$ does not exist, causing larger errors.
\revision{
Obviously, arbitrary scale-free networks that may exhibit strong structural inhomogeneities---for instance, networks consisting of only sparsely connected communities---would require additional variables to retain accuracy, such as separate weighted shares for each community. Low-dimensional approximations for these types of networks could be investigated in future work.
}

The inspiration for the mMFE~\eqref{eq:modified_mfe} arose from findings in~\cite{Luecke2024}, where a computational approach was proposed to identify forecastable variables of dynamical processes on networks.
For such processes, reduced modeling is typically guided by theoretical insights---our approach therefore constitutes an exception.
Along these lines, it would be highly desirable to gain a theoretical understanding of the observed superior behavior of the mMFE.
Furthermore, this work also highlights the importance of data-driven approaches, since many of the insights presented here were supported by computational analysis.
These methods could be utilized in future work to derive low-dimensional approximations for other systems that are not yet well understood analytically.

This work in particular could also be extended by refining the mMFE in parameter regimes where its accuracy is limited, for instance by allowing a time-dependent $\ifac$. Moreover, a modified version of the PA that operates on the degree-weighted shares $\wshv$ could be explored.

\begin{acknowledgments}
This work has been funded by Deutsche Forschungsgemeinschaft (DFG, German
Research Foundation) -- project ID: 546032594.
\end{acknowledgments}

\appendix
\renewcommand{\thefigure}{S\arabic{figure}}
\setcounter{figure}{0}

\section{Derivation of mMFE for more than two opinions} \label{sec:appendix}

Recall that in the case of $\numO > 2$ opinions, the degree-weighted shares $\wshv \in [0,1]^\numO$ are defined via $\wshv = (\wsh_\oip)_{\oip \in [\numO]}$,
\begin{equation}
	\wsh_\oip :=
	\frac{1}{\sum_\nip d_\nip} \sum_{\nip\in[\numA]} d_\nip  \E[\indx{\nip}{\oip}],
\end{equation}
where $\indx{\nip}{\oip} \in \{0,1\}$ denotes the indicator function of the event $\x_\nip = \oip$.
The rate at which node $\nip$ in state $\x_\nip = \oip$ transitions to a different state $\ois$ is given by
\begin{equation}
	r_{\oip,\ois} \frac{1}{d_\nip} \sum_{ \nis \in [\numA] }  A_{\nip, \nis} \indx{\nis}{\ois} + \tilde{r}_{\oip, \ois} =: q_{\oip, \ois}^\nip.
\end{equation}
Hence, for all $\oip \in [\numO]$ it is
\begin{align}
	\frac{\dif}{\dif t} \E[\indx{\nip}{\oip}]
	 & = \sum_{\ois \neq \oip} \P(\x_\nip = \ois)\ \E\big[q_{\ois, \oip}^\nip \mid \x_\nip = \ois\big]    \\
	 & \ - \sum_{\ois \neq \oip} \P(\x_\nip = \oip)\ \E\big[q_{\oip, \ois}^\nip \mid \x_\nip = \oip\big].
\end{align}
Using the equality $\P(\x_\nip = \ois)\ \E[\indx{\nis}{\oip} \mid \x_\nip = \ois] = \E[\indx{\nis}{\oip} \indx{\nip}{\ois}]$, it follows that
\begin{align}
	 & \frac{\dif}{\dif t} \E[\indx{\nip}{\oip}]                                                                                                                                          \\
	 & = \sum_{\ois \neq \oip}\ \frac{r_{\ois,\oip}}{d_\nip} \sum_\nis A_{\nip, \nis} \E[\indx{\nip}{\ois} \indx{\nis}{\oip}] + \tilde{r}_{\ois, \oip} \E[\indx{\nip}{\ois}] \nonumber    \\
	 & \ - \sum_{\ois \neq \oip}\ \frac{r_{\oip,\ois}}{d_\nip} \sum_\nis A_{\nip, \nis} \E[\indx{\nip}{\oip} \indx{\nis}{\ois}] - \tilde{r}_{\oip, \ois} \E[\indx{\nip}{\oip}]. \nonumber
\end{align}
Thus, we have
\begin{align}
	 & \frac{\dif}{\dif t} \wsh_\oip = \frac{1}{\sum_\nip d_\nip} \sum_\nip d_\nip \frac{\dif}{\dif t} \E[\indx{\nip}{\oip}] = \frac{1}{\sum_\nip d_\nip} \sum_{\ois \neq \oip} \Big( \\
	 & r_{\ois,\oip} \sum_{\nip,\nis} A_{\nip, \nis} \E[\indx{\nip}{\ois} \indx{\nis}{\oip}] + \tilde{r}_{\ois, \oip} \sum_\nip d_\nip \E[\indx{\nip}{\ois}] -\nonumber               \\
	 & r_{\oip,\ois} \sum_{\nip,\nis} A_{\nip, \nis} \E[\indx{\nip}{\oip} \indx{\nis}{\ois}] - \tilde{r}_{\oip, \ois} \sum_\nip d_\nip \E[\indx{\nip}{\oip}] \Big). \nonumber
\end{align}
Next, we apply the \emph{(weak) correlation assumption}
\begin{multline}
	\sum_{\nip,\nis} A_{\nip, \nis} \E[\indx{\nip}{\ois} \indx{\nis}{\oip}] \overset{!}{=}  \\
	\sum_{\nip,\nis} A_{\nip, \nis}\, \ifac\, \E[\indx{\nip}{\ois}] \E[\indx{\nis}{\oip}]
\end{multline}
and the \emph{(weak) interchangeability assumption}
\begin{equation}
	\sum_{\nip,\nis} A_{\nip, \nis} \E[\indx{\nip}{\ois}] \E[\indx{\nis}{\oip}] \overset{!}{=} \sum_{\nip,\nis} A_{\nip, \nis}\, \wsh_\ois\, \wsh_\oip
\end{equation}
for all $\oip \neq \ois$. This yields
\begin{align}
	\frac{\dif}{\dif t} \wsh_\oip = \sum_{\ois \neq \oip}\Big( & \ifac\ r_{\ois,\oip}\ \wsh_\ois\, \wsh_\oip + \tilde{r}_{\ois,\oip}\, \wsh_\ois         \\
	                                                           & - \ifac\ r_{\oip,\ois}\ \wsh_\oip\, \wsh_\ois - \tilde{r}_{\oip,\ois}\, \wsh_\oip \Big)
\end{align}
for all $\oip \in [M]$,
which is equivalent to the mMFE \eqref{eq:modified_mfe} when written as a system of equations.

\section{\revision{Scale-free networks generated by the configuration model}} \label{appendix:configuration-model}
\begin{figure*}
	\centering
	\includegraphics[width=.9\textwidth]{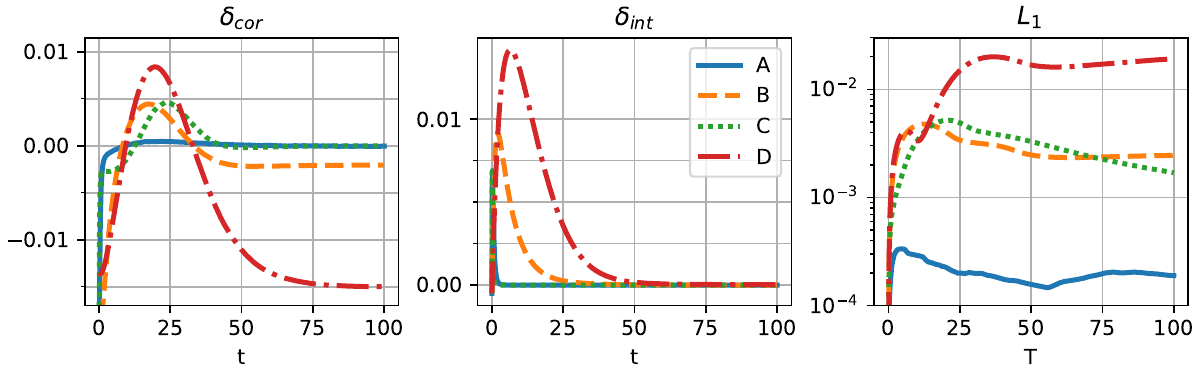}
	\vspace{-.3cm}
	\caption{Error analysis for the configuration model, cf. Fig. \ref{fig:errors_assumptions} in the main text.}
	\label{fig:errors_assumptions_conf}
\end{figure*}
In this section we verify that the mMFE \eqref{eq:modified_mfe} is also valid for scale-free networks generated by the \textit{configuration model} \cite{Bollobas1980}, which can be used to generate a random graph with an arbitrary degree sequence $(d_\nip)_{\nip \in [\numA]}$.
The method works as follows: After assigning $d_\nip$ \textit{half-edges} to each node $\nip$, two half-edges are repeatedly picked from all remaining half-edges, joined together to form an edge, and removed from the set of available half-edges, until all half-edges have been used up.
This potentially yields a \textit{multigraph} since self-loops and multiple edges between two nodes are possible.
In such a case we simply restart and repeat the whole procedure until finding a simple graph.
To produce a scale-free network, we first generate a degree sequence $(d_\nip)_{\nip \in [\numA]}$ such that the associated degree distribution follows a power law with exponent $\gamma=3$, and then apply the configuration model.
To facilitate a fair comparison to a BA model with preferential attachment parameter $\bap$, the minimum degree in the degree sequence should also be chosen as $\bap$.

Asymptotically, the BA model and the configuration model are relatively similar.
For example, for both models the average clustering coefficient falls to zero rapidly at a rate of approximately $1 / N$, and the diameter grows moderately at a rate of approximately $\log N$ \cite{Luecke2024a}.
Furthermore, the assortativity, i.e., the degree correlation between neighbors, is asymptotically zero for both models.
However, for the finite network sizes of $\numA \approx 10^4$ considered here, the BA model and the configuration model exhibit significant differences.
The most noteworthy aspect is that for the BA model the assortativity is still clearly negative, meaning that it is more likely for small degree nodes to be connected to large degree nodes, while for the configuration model the assortativity is already close to zero.
Moreover, the BA model shows a considerably larger clustering coefficient than the configuration model.

\textbf{Results.}
We conducted the same error and robustness analysis for the configuration model as presented for the BA model in the main text.
The error analysis in the four regimes A, B, C, D is shown in Fig. \ref{fig:errors_assumptions_conf} and is almost identical to the BA model.
The only significant difference is that for the configuration model the errors in the noise-less regime $C$ are lower.
This is likely caused by the aforementioned lack of clustering and assortativity compared to the BA model, which hinders the formation of clusters of identical opinion.

The robustness analysis for the configuration model is shown in Fig. \ref{fig:errors_robust_conf} and is also very similar to the BA model.
The maximum error and the variance of errors is slightly larger for the configuration model, but the median error is also below 0.003.

The optimal correlation factor for the configuration model (with minimum degree 3) is $\ifac = 0.75$, which is slightly smaller than the factor $\ifac = 0.81$ determined for the BA model with preferential attachment parameter $\bap=3$ in the main text.
This difference could be explained by the negative assortativity of the BA model which implies that small degree nodes are more likely connected to large degree nodes.
Since such a pair of nodes with small and large degree is less correlated than a pair of nodes with small degrees, the correlation factor $\ifac$ is smaller for the configuration model (indicating a larger positive correlation of neighbors).
A more thorough investigation could be conducted in future work.

\begin{figure}[h!]
	\centering
	\includegraphics[width=\linewidth]{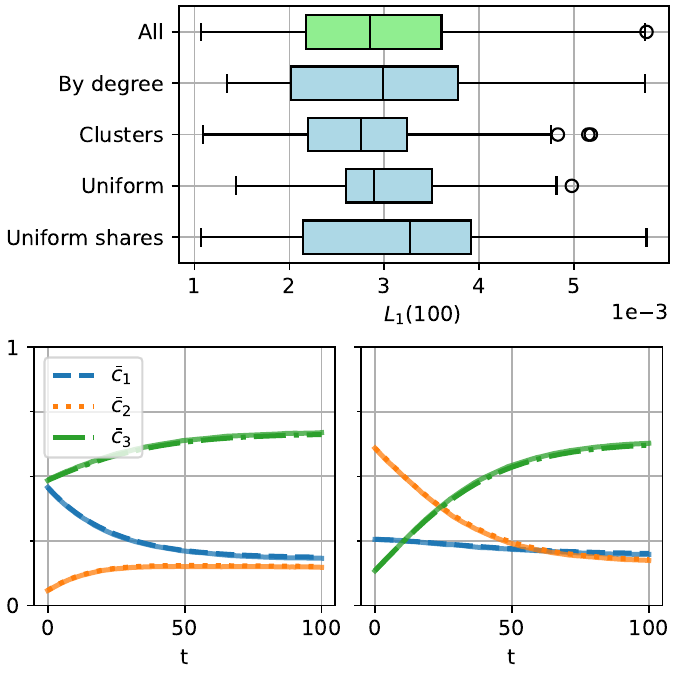}
	\vspace{-0.7cm}
	\caption{Robustness of the mMFE for the configuration model, cf. Fig. \ref{fig:errors_robust} in the main text.}
	\label{fig:errors_robust_conf}
\end{figure}

\section{\revision{Pair-approximation}} \label{appendix:pair-approx}
In this section we briefly outline the derivation of the classical \emph{pair-approximation} in the case of $\numO=2$ opinions.
A more detailed derivation can readily be found in the literature, see for instance \cite{Peralta2018, Luecke2024a}.
Instead of closing the first-order moment equation using an independence assumption $\E[(1-x_\nip)x_\nis] \approx \E[1-x_\nip] \E[x_\nis]$, the pair approximation assumes that all neighbor pairs are statistically 
equivalent and introduces the average pair probability
$\mathbb{E}[(1-x_i)x_j] =: s$ as an additional variable. Based on \eqref{eq:E}, this yields
\begin{align}
	\frac{\dif}{\dif t} \sh = s (r_{0,1} - r_{1,0}) -\sh (\tilde{r}_{0,1} + \tilde{r}_{1,0}) + \tilde{r}_{0, 1}.
\end{align}
To obtain the evolution equation of $s$, we then have to consider the second-order moment equation, which in turn contains third-order moments, e.g., terms of the form $\E[\x_\nit ( 1 - \x_\nip ) \x_\nis]$ that describe the probability of a connected triplet of nodes $\nit - \nip - \nis$ being in states $(1, 0, 1)$.
We close this second-order moment equation by assuming that the outer two nodes $\nit$ and $\nis$ of the triplet are conditionally independent given $x_\nip$, which yields
\begin{align}
	\E[\x_\nit ( 1 - \x_\nip ) \x_\nis] &\approx \frac{\E[(1 - \x_\nip) \x_\nit]}{1 - \E[\x_\nip]} \E[(1 - \x_\nip) \x_\nis] \\
    &\approx \frac{s^2}{1 - \sh}.
\end{align}
Following this procedure for all occurring third-order moments yields this two-dimensional system of equations called \emph{pair-approximation}:
\begin{align}
	\frac{\dif}{\dif t} \sh & = s (r_{0,1} - r_{1,0}) -\sh (\tilde{r}_{0,1} + \tilde{r}_{1,0}) + \tilde{r}_{0, 1}.                        \\
	\frac{\dif}{\dif t} s  & = -2 \frac{d-1}{d} \frac{s^2}{1-\sh} r_{0,1} -2 \frac{d-1}{d} \frac{s^2}{\sh} r_{1,0}                      \\
	                        & \quad + s \Big(\frac{d-2}{d} r_{0,1} + \frac{d-2}{d} r_{1,0} - 2 \tilde{r}_{0,1} - 2 \tilde{r}_{1,0}  \Big) \\
	                        & \quad + \sh \Big( \tilde{r}_{1,0} -\tilde{r}_{0,1} \Big) + \tilde{r}_{0, 1},
\end{align}
where $d$ denotes the average degree in the graph.

\section{Supplementary figures}
In this section, two supplementary figures are provided.
Figure \ref{fig:error_vs_rates} shows the dependence of the mMFE error on the
difference of imitation rates, and Fig. \ref{fig:network_size} shows the mMFE
error for varying numbers of nodes.

\begin{figure}[h!]
	\centering
    \includegraphics[width=.9\linewidth]{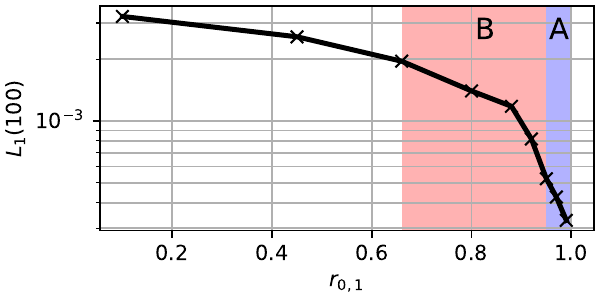}
	\vspace{-0.3cm}
	\caption{Error of the mMFE for the parameters $r_{1,0} = 1$, $\tilde{r}_{1,0}=\tilde{r}_{0,1}=0.01$ and varying $r_{0,1}$. The correlation factor $\ifac$ was optimized for each system, resulting in values between $0.8$ and $0.84$. The plot also shows the regions of regimes A and B as defined in the main text.}
	\label{fig:error_vs_rates}
\end{figure}

\begin{figure}[h!]
	\centering
    \includegraphics[width=.9\linewidth]{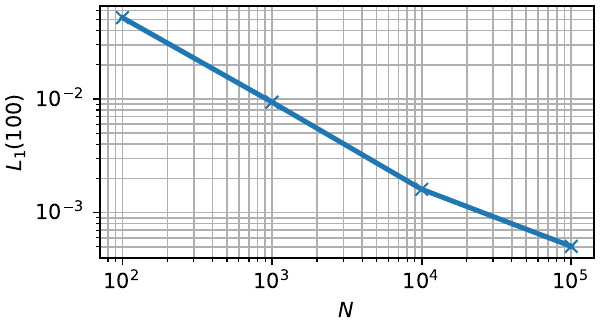}
	\vspace{-0.3cm}
	\caption{Error of the mMFE for the parameters $r_{0,1} = 1$, $r_{1,0} = 0.98$, $\tilde{r}_{0,1}=\tilde{r}_{1,0}=0.01$  and $\ifac = 0.81$ against the number of nodes $\numA$ of the BA network.}
	\label{fig:network_size}
\end{figure}

\bibliography{references.bib}

\end{document}